\documentstyle[aps,prl,floats,epsfig]{revtex}

\newcommand{\Exp}[1]{{\mathrm{e}}^{\mbox{\footnotesize$#1$}}}
\newcommand{\power}[1]{^{\mbox{\footnotesize$#1$}}}
\newcommand{\rewop}[1]{_{\mbox{\footnotesize$#1$}}}
\newcommand{\tfrac}[2]{{\textstyle\frac{#1}{#2}}}

\begin{document}
\draft

\title{Separability of Two-Party Gaussian States}

\author{Berthold-Georg Englert$^{\dag}$ and
Krzysztof W\'odkiewicz$^{\ddag,*}$ }
\address{$^\dag$Max-Planck-Institut f\"ur Quantenoptik, %
Hans-Kopfermann-Stra\ss{}e 1, 85748 Garching, Germany\\%
$^\ddag$Instytut Fizyki Teoretycznej, Uniwersytet Warszawski, %
ul.\ Ho\.za 69, Warszawa 00--681, Poland\\%
$^{*}$Department of Physics and Astronomy, %
University of New Mexico, Albuquerque NM 87131}

\date{Brief Report submitted to Physical Review A, received 26 July 2001}

\ifpreprintsty\relax\else\wideabs{\fi
\maketitle

\begin{abstract}%
We investigate the separability properties of quantum two-party Gaussian
states in the framework of the operator formalism for the density operator.
Such states arise as natural generalizations of the entangled state originally
introduced by Einstein, Podolsky, and Rosen. 
We present explicit forms of separable and nonseparable Gaussian states.
\ifpreprintsty
\end{abstract}
\else\\{}
\end{abstract}}
\fi

\narrowtext

\section{Introduction}
In their reasoning concerning the alleged incompleteness 
of quantum mechanics \cite{epr},
Einstein, Podolsky, and Rosen (EPR) used this wave function for a 
system composed of two particles:
\begin{equation}
\label{epr} \Psi (x_{1},x_{2})= \int_{-\infty}^{\infty}dp\, 
\Exp{(2\pi i/h)(x_1-x_2+x_0)p}\;. 
\end{equation}
It is a singular function of the distance $x_1-x_2$ and could be visualized 
as an infinitely sharp Gaussian wave function of the entangled two-party 
system. 
Bell inequalities of some kind are violated for this wave function, as can be
demonstrated by using its Wigner representation~\cite{kbkw}. 

Recent applications of entangled two-mode squeezed states of light for 
quantum teleportation \cite{fsbfkp} and other quantum information purposes
\cite{opkp} have generated a lot of interest in the separability properties 
of general mixed Gaussian states in quantum optics \cite{kimble}. 
In one approach, the separability properties of continuous-variable systems 
in states described by Gaussian Wigner functions have been investigated
with the aid of Heisenberg uncertainty relations \cite{zoller}. 
Another approach made use of the criterion of positivity under partial
transposition \cite{simon}. 
Both approaches use the basic definition, namely that a general 
quantum density operator of a two-party system is separable if it is a convex
sum of product states \cite{werner}:
\begin{equation}
\label{separability}
 \rho = \sum_k p_k\,\rho_a^{(k)}\otimes\rho_b^{(k)}\quad
\text{with $\sum_k p_k=1$ and $p_k>0$}\,, 
\end{equation}
where $\rho_a^{(k)}$ and $\rho_b^{(k)}$ are statistical operators of the two
subsystems in question. 

The authors of \cite{zoller} and \cite{simon} arrived at essentially the same
conclusions, while using very different techniques. 
The objective of this Brief Report is to show how equivalent results are
derived by employing the powerful algebraic methods of quantum optics.
We use a direct operator method to study the
separability of an arbitrary two-party Gaussian operator
$G(a,a^{\dagger},b,b^{\dagger})$ of unit trace,
referring, for instance, to two modes of the radiation field, 
parameterized by their ladder operators $a$ and $b$.
An explicit algebraic form of the Gaussian operator enables us to decide 
whether $G$ is a density operator and, if so, whether it is separable, 
that is: whether $G$ is a positive operator of the form (\ref{separability}). 
Our method works for arbitrary Gaussian operators \cite{tutorial}, 
but for the sake of clarity and also in view of the
importance of the EPR wave function (\ref{epr}), here we shall carry out 
the explicit calculations only for a specific class of Gaussian operators 
that form a natural generalization of the original EPR state (\ref{epr}). 
We wish, however, to stress that the algebraic approach is quite general, 
and that the method provides an explicit construction of the Werner
decomposition (\ref{separability}) for separable two-party Gaussian states. 
The algebraic approach provides a natural link between the
partial-transposition criterion of Peres \cite{peres} and $P$-representable
Gaussian operators.

\section{Gaussian operators. Basics}
Following Wigner \cite{wigner}, we associate a real phase space function
$W(\alpha,\beta)$ with any such operator, and in particular with the Gaussian
operators $G$ of interest, in accordance with \cite{cahglau}
\begin{eqnarray}
\label{defgauss}
G &=& 2^{2} \int\! d^{2}\alpha \int\! d^{2}\beta\,W(\alpha,\beta)\ 
\nonumber\\&&
\times
: \Exp{-2(a^{\dagger}-\alpha^*)(a-\alpha)}
\Exp{-2(b^{\dagger}-\beta^*)(b-\beta)} :\,.
\end{eqnarray}
Here, the integrations are over the two phase spaces of the oscillators,
parameterized by the complex variables $\alpha$ and $\beta$,
and the normally ordered exponential operators 
${:\exp\bigl(-2(a^{\dagger}-\alpha^*)(a-\alpha)\bigr):}$\,, 
${:\exp\bigl(-2(b^{\dagger}-\beta^*)(b-\beta)\bigr):}$ 
are  parity operators \cite{wig+par},
\begin{equation}
  \label{parops}
  (-1)\power{a^{\dagger}a}=\ :\Exp{-2a^{\dagger}a}:\;,\quad
  (-1)\power{b^{\dagger}b}=\ :\Exp{-2b^{\dagger}b}:\;,
\end{equation}
displaced in phase space by $\alpha$ and $\beta$. 
We take for granted that the traces of $aG$ and $bG$ vanish; otherwise a
unitary linear transformation would enforce this condition.

A Gaussian operator $G$, then, is one whose Wigner function $W$ is a Gaussian 
function of its complex variables $\alpha$ and $\beta$:
\begin{equation}
  \label{gausswig}
  W(\alpha,\beta)=\frac{\sqrt{ \det \mathbf{W}}}{\pi^{2}}\
   \Exp{-\frac{1}{2}{\mathbf{v}}^{\dagger}{\mathbf{W}}{\mathbf{v}}}\,,
\end{equation}
where $ \mathbf{v}^{\dagger}=[\alpha^*,\alpha,\beta^*,\beta]$ 
is a complex 4-component row and
${\mathbf{W}}>0$ is a positive $4\times4$-matrix.
The positivity of $\mathbf{W}$ and the prefactor in (\ref{gausswig})
ensure the correct normalization of $G$ to unit trace,
\begin{equation}
\label{normalization} 
\mathop{\mathrm{Tr}} G
=\int\! d^{2} \alpha \int\! d^{2} \beta\, W(\alpha,\beta)=1\,.
\end{equation}

We shall also find it useful to work with the Weyl-Wigner characteristic
function
\begin{equation}\label{defC}
C(\alpha,\beta)=\mathop{\mathrm{Tr}}\left\{
\Exp{\alpha a^{\dagger}-\alpha^*a}\,G\,\Exp{\beta b^{\dagger}-\beta^*b}\right\}
= \Exp{-\frac{1}{2}\mathbf{v}^{\dagger}\mathbf{V}\mathbf{v}}\,,
\end{equation}
a Gaussian function as well,
related to the Wigner function $W(\alpha,\beta)$ by two-fold
complex Fourier transformation.
Accordingly, we have
\begin{equation}
  \label{W->V}
  {\mathbf{V}} = {\mathbf{E}\mathbf{W}}^{-1}\mathbf{E}>0
\quad\text{and}\quad {\mathbf{W}} = {\mathbf{E}\mathbf{V}}^{-1}\mathbf{E}\,,
\end{equation}
where ${\mathbf{E}}= \text{diag}[1,-1,1,-1]$ is a diagonal $4\times4$ matrix.

Given $W(\alpha,\beta)$ and $C(\alpha,\beta)$, Eqs.\ 
(\ref{gausswig}) and (\ref{defC}) do not specify
$\mathbf{W}$ and $\mathbf{V}$ uniquely.
The symmetry of our standard form of $\mathbf{V}$, 
\begin{equation}
  \label{standardV}
{\mathbf{V}} 
= \left[ {\begin{array}{cccc} 
n_1+\frac{1}{2} & m_1 & m_s & m_c \\ 
m_1^* & n_1+\frac{1}{2}& m_c^* & m_s^* \\ 
m_s^* & m_c & n_2+\frac{1}{2} & m_2 \\ 
m_c^* & m_s & m_2^* & n_2+\frac{1}{2}
\end{array}}
\right]
\end{equation}
with real $n_1$, $n_2$ and complex $m_1$, $m_2$, $m_s$, $m_c$,
exploits this arbitrariness conveniently.

\section{Gaussian operators. Explicit forms}
As stated in the Introduction, we shall illustrate the algebraic method
by the example of a generalized version of the EPR wave function
(\ref{epr}). 
This generalized Gaussian operator is specified by a $\mathbf{V}$ matrix of
the restricted form
\begin{equation} 
{\mathbf{V}} 
= \left[ {\begin{array}{cccc} n+\frac{1}{2} & 0 & 0 & m \\ 0 &
n+\frac{1}{2}& m^* & 0 \\ 0 & m & n+\frac{1}{2} & 0 \\ m^* & 0 & 0
& n+\frac{1}{2}
\end{array}}
\right]\;, 
\label{thisV}
\end{equation}
corresponding to ${n_1=n_2=n}$, ${m_1=m_2=m_s=0}$, and ${m_c=m}$ in
(\ref{standardV}). 
The constraint
\begin{equation}\label{constraint0}
n+\tfrac{1}{2}> |m|
\end{equation} 
ensures ${\mathbf{V}}>0$.
The significance of $n$ and $m$,
\begin{equation}
  \label{nm=Tr}
  n=\mathop{\mathrm{Tr}}\bigl\{a^{\dagger}a\,G\bigr\}
   =\mathop{\mathrm{Tr}}\bigl\{b^{\dagger}b\,G\bigr\}\,,\quad
  m=-\mathop{\mathrm{Tr}}\bigl\{ab\,G\bigr\}\,,
\end{equation}
is revealed upon expanding (\ref{defC}) in powers of $\mathbf{v}$.

In view of (\ref{W->V}), the matrix $\mathbf{W}$ of the Wigner function
(\ref{gausswig}) is at hand, and then (\ref{defgauss}) gives us the operator
in normally ordered form.
With the identity \cite{schwinger}
\begin{equation}
  \label{:exp:}
  :\Exp{-\zeta a^{\dagger}a}:\,=(1-\zeta)\power{a^{\dagger}a}\,, 
\end{equation}
valid for all complex $\zeta$ (the $\zeta=2$ case is met in (\ref{parops})), 
we so arrive at one
explicit form of the corresponding Gaussian operator $G$, namely
\begin{eqnarray}
\label{explicitgauss}
 G&=&\frac{1}{(n+1)^2-|m|^2}\,
\Exp{-\frac{m}{(n+1)^2-|m|^2}a^{\dagger}b^{\dagger}}
\nonumber
\\&&\times
\left[\frac{n(n+1)-|m|^2}{(n+1)^2-|m|^2}\right]
\power{a^{\dagger}a+b^{\dagger}b}
\Exp{-\frac{m^*}{(n+1)^2-|m|^2}a b}\,.
\end{eqnarray}
More compactly, this appears as
\begin{equation}
\label{sandwich} 
G =S^{\dagger}\, G_1\, G_2\,S\,,
\end{equation}
where the basic Gaussian operators $G_1$, $G_2$ have the form
\begin{equation}
\label{basicgauss} 
G_1 = (1-g_1)\,g_1\power{a^{\dagger}a} \quad
\text{and} \quad 
G_2 = (1-g_2)\,g_2\power{b^{\dagger}b}
\end{equation}
with
\begin{equation}
  \label{ga+gb}
  g_1=g_2=\frac{n(n+1)-|m|^2}{(n+1)^2-|m|^2}\,,
\end{equation}
and the sandwiching operator $S$ is 
\begin{equation}
  \label{defS}
  S=\frac{\sqrt{(n+1)^2-|m|^2}}{n+1}\,\Exp{-\frac{m^*}{(n+1)^2-|m|^2}a b}\,.
\end{equation}

Quite generally, the basic Gaussians of (\ref{basicgauss}) have a finite trace
if $-1\leq g_1,g_2\leq1$. 
More specifically, the constraint (\ref{constraint0}) implies here that
$-(4n+3)^{-1}<g_1=g_2<1$ and thus ensures the finite value of the trace.
Positivity of $G_1$ and $G_2$, and therefore also of $G$ itself, requires more
restrictively that $0\leq g_1,g_2\leq1$ in (\ref{basicgauss}), irrespective of
the particular form that $S$ might have.
For $\mathbf{V}$ of the specific form (\ref{thisV}), this says that $G>0$ is
equivalent to
\begin{equation} \label{constraint1}
n(n+1)\geq |m|^2\quad\mbox{or}\quad
n\geq\sqrt{|m|^2+\tfrac{1}{4}\,}-\tfrac{1}{2}\,.
\end{equation}
As a compact statement about $\mathbf{V}$, this appears as
\begin{equation}
  \label{constraint1'}
  {\mathbf{V}}+\tfrac{1}{2}{\mathbf{E}}\geq0
\end{equation}
with the diagonal matrix $\mathbf{E}$ of (\ref{W->V}).

Note that, if (\ref{constraint0}) is obeyed but (\ref{constraint1}) is not,
then we have a Gaussian operator that does not represent a density operator
although its Wigner function is positive and properly normalized because
the matrices $\mathbf{V}$ and $\mathbf{W}$ are positive. 
These matters are illustrated in Fig.~\ref{TheFig}.
 
\ifpreprintsty\relax\else
\begin{figure}[!t]
\centerline{\epsfig{file=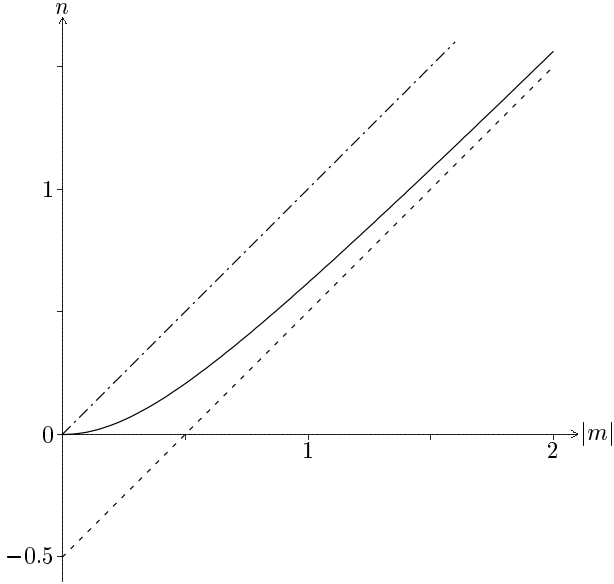,width=0.95\columnwidth}} 
\caption{\label{TheFig}%
Concerning the parameters of the Gaussian operator associated with the
$\mathbf{V}$ matrix of (\ref{thisV}).
Only $n,|m|$ values above the dashed line are allowed by constraint
(\ref{constraint0}). 
According to (\ref{constraint1}), values on or above the solid line specify
positive Gaussians of the form (\ref{pureG}). 
For values on the solid line, the Gaussian operator is a projector.
Separable Gaussians belong to values on or above the dash-dotted line;
see (\ref{cond:Prepr}).}
\end{figure}
\fi

The explicit construction of the Gaussian operator was here performed 
for a matrix $\mathbf{V}$ of the specific form (\ref{thisV}). 
In the most general situation of (\ref{standardV}), 
we have more parameters, but $G$ is always of the generic form 
(\ref{sandwich}), that is: a product $G_1G_2$ of two thermal Gaussian 
operators, sandwiched by a $S^{\dagger}$, $S$ pair.

\section{Gaussian operators. Pure states}\label{sec:pure}
In this section we investigate Gaussian operators that are projectors
and thus represent pure states. 
This case occurs when the equal sign holds in (\ref{constraint1}), so that
$g_1=g_2=0$, and
\begin{equation}
  \label{g->0}
  (1-g_1)\,g_1\power{a^{\dagger}a}\to0\power{a^{\dagger}a}
  =\delta\rewop{a^{\dagger}a,0} 
  \quad\text{as $g_1\to0$},
\end{equation}
for example, states that
\begin{equation}
  \label{g->0'}
  G_1G_2\to 0\power{a^{\dagger}a+b^{\dagger}b}=
 \delta\rewop{a^{\dagger}a,0}\delta\rewop{b^{\dagger}b,0}=
 |0,0\rangle\langle0,0|
\end{equation}
in this limit.
The Hilbert space vector $|n,m\rangle$, here for $n=m=0$, denotes the state
with $n$ quanta of $a$-type and $m$ of $b$-type.

As a consequence, Eq.\ (\ref{explicitgauss}) turns into
\begin{equation}
\label{pureG}
 G=|\Psi\rangle\langle\Psi|
=\bigl(1-|\lambda|^2\bigr) \Exp{\lambda a^{\dagger}b^{\dagger}} 
|0,0\rangle\langle0,0| \Exp{\lambda^* ab}
\end{equation} 
where $\lambda=-m/(n+1)$. 
Such an operator projects onto
\begin{equation}
\label{purestate}
 |\Psi\rangle=\sqrt{1-|\lambda|^{2}}
\sum_{n=0}^{\infty}\lambda^{n}\ |n,n\rangle\,.
\end{equation}
For real $\lambda$, we recognize an example of 
the well known two-mode squeezed state that can be
generated by Nondegenerate Optical Parametric Amplification,
\begin{equation}  
|{\mathrm{NOPA}}\rangle = \Exp{r(a^{\dagger}b^{\dagger}-ab)}|0,0\rangle 
= \frac{1}{\cosh r} \sum_{n=0}^{\infty} (\tanh r)^{n} |n,n\rangle\,,
\end{equation}
where $\lambda=\tanh r$ relates the squeezing parameter $r$ to $\lambda$ and
thus to parameter $n$ of the Gaussian operator.

We said above that (\ref{explicitgauss}) is a natural generalization of the
EPR state (\ref{epr}).
The stage is now set for justifying this remark.
To this end, we first note that, for real $\lambda$, 
the position wave function of $|\Psi\rangle$ of (\ref{purestate}) is given by
\begin{eqnarray}
\Psi (x_{1},x_{2}) &=& \frac{ 1}{\sqrt{\pi}} \exp\left(-\frac{(1+\lambda^{2})
(x_{1}^{2}+ x_{2}^{2}) -4\lambda x_{1}x_{2}} { 2(1-\lambda^{2})}\right)
\nonumber\\
&=&\frac{1}{\pi\hbar}\int\!dp\,\Exp{(i/\hbar)(x_1-x_2)p}
\nonumber\\&&\times
\sqrt{\frac{1-\lambda}{1+\lambda}}
\Exp{-\frac{1-\lambda}{1+\lambda}[(p/\hbar)^2+\frac{1}{4}(x_1+x_2)^2]}\,,
\end{eqnarray}
and then observe that
\begin{equation}
\Psi (x_{1},x_{2})\propto\int\!dp\,\Exp{(i/\hbar)(x_1-x_2)p}
\end{equation}
obtains in the limit $\lambda\to1$.
Indeed, this is the EPR state of (\ref{epr}) with $x_0=0$.

\section{Separability of Gaussian states}
A positive Gaussian operator $G$ is said to be $P$-re\-pre\-sent\-able 
if it can be written in the following form:
\begin{eqnarray}
G&=&  \int\! d^{2}\alpha \int\! d^{2}\beta\,P(\alpha,\beta)
\nonumber\\&&\quad\times 
:\Exp{-(a^{\dagger}-\alpha^*)(a-\alpha)}
\Exp{-(b^{\dagger}-\beta^*)(b-\beta)}:\,,
\label{Prep}
\end{eqnarray}
with a non-negative phase-space function
$P(\alpha,\beta)$ that must not be more singular than a
Dirac $\delta$ function.
The  ordered exponentials 
${:\exp\bigl(-(a^{\dagger}-\alpha^*)(a-\alpha)\bigr):}$ and
${:\exp\bigl(-(b^{\dagger}-\beta^*)(b-\beta)\bigr):}$ 
are projectors onto the coherent states labeled by $\alpha$ and $\beta$,
respectively.  

For a $P$-representable Gaussian operator, we have
\begin{equation}
  \label{gaussp}
  P(\alpha,\beta)=\frac{\sqrt{ \det {\mathbf{P}}}}{\pi^{2}}\
   \Exp{-\frac{1}{2}{\mathbf{v}}^{\dagger}{\mathbf{P}}{\mathbf{v}}}
\end{equation}
with $\mathbf{P}$ related to the $4\times4$ matrix $\mathbf{V}$ of the
characteristic function (\ref{defC}) and the matrix $\mathbf{W}$ of the
Wigner function (\ref{gausswig}) by 
\begin{equation}
{\mathbf{P}} 
={\mathbf{E}}\bigl({\mathbf{V}}- \tfrac{1}{2} {\mathbf{I}}\bigr)^{-1} 
{\mathbf{E}} =\bigl({\mathbf{W}}^{-1}- \tfrac{1}{2} {\mathbf{I}}\bigr)^{-1} 
\,,
\end{equation}
where $\mathbf{I}$ is the $4\times4$ unit matrix.
So, a given Gaussian operator is $P$-representable if
\begin{equation}
{\mathbf{V}}- \tfrac{1}{2} {\mathbf{I}}\geq0\;.
\end{equation}
If the left-hand side is truly positive, 
we have a  four-dimensional Gaussian in (\ref{gaussp}), 
else it is the product of a two-dimensional Gaussian and a
two-dimensional $\delta$ function, or the product of two two-dimensional
$\delta$ functions.

For the Gaussian operator (\ref{explicitgauss}), the existence of the
$P$-representation is guaranteed if
\begin{equation}
n \geq |m|\,.
\label{cond:Prepr}
\end{equation}
In Fig.\ \ref{TheFig} these $n,|m|$ values are on or above the dash-dotted
line. 
Since the ordered exponentials in (\ref{Prep}) project to coherent states,
such $P$-representable Gaussians are convex sums of product states.
They are thus of the separable kind, as defined in (\ref{separability}),
where the formal summation over $k$ is now the two-fold phase-space
integration of (\ref{Prep}).

As Peres observed \cite{peres}, the partial transpose $\rho^{T_a}$ of a
separable statistical operator (\ref{separability}) is another statistical
operator because the $\rho_a^{(k)}$'s are turned into other positive operators
and the $\rho_b^{(k)}$'s are not affected to begin with.
In other words, ${\rho^{T_a}\geq0}$ is a \emph{necessary} property of a
separable $\rho$.
As surmised by Peres and demonstrated by the Horodecki family
\cite{horodeccy}, it is in fact \emph{sufficient} for two-party systems
composed of two spin-$\frac{1}{2}$ objects (``qubits'') or of one qubit and
one spin-$1$ object.

Concerning the systems of interest here, of two harmonic oscillators, the
Peres criterion  ${\rho^{T_a}\geq0}$ does not imply that $\rho$ is separable.
But, as Simon noted \cite{simon}, 
in the particular case that $\rho$ is a positive Gaussian operator, Peres'
criterion \emph{is} sufficient to ensure that $\rho$ is separable.  
Indeed, for a positive Gaussian $G$ the Peres criterion is equivalent to
requiring that $G$ is $P$-representable. 

To see this, let us be more specific and agree on using the Fock
representation to define the partial transpose.
Then, ${G\to G^{T_a}}$ amounts to ${W(\alpha,\beta)\to W(\alpha^*,\beta)}$ in
(\ref{defgauss}), that is:
\begin{equation}
  \label{wigTa}
  {\mathbf{W}}\to{\mathbf{T}}_a{\mathbf{W}}{\mathbf{T}}_a
\quad\mbox{with}\quad
{\mathbf{T}}_a=\left[
\begin{array}{cccc}
0&1&0&0 \\ 1&0&0&0 \\ 0&0&1&0 \\ 0&0&0&1
\end{array}
\right]
\end{equation}
in (\ref{gausswig}) and, as (\ref{W->V}) implies, 
${\mathbf{V}}\to{\mathbf{E}}{\mathbf{T}}_a{\mathbf{E}}\,{\mathbf{V}}\,
{\mathbf{E}}{\mathbf{T}}_a{\mathbf{E}}$
in (\ref{defC}).

For the $\mathbf{V}$ matrix of (\ref{thisV}), this results in
\begin{eqnarray}
G^{T_{a}}&=&\frac{1}{n+1+|m|} \left(\frac{n+|m|}{n+1+|m|}\right)
\power{\frac{1}{2}(a^{\dagger}-\mu b^{\dagger})(a-\mu^* b)}
\nonumber\\&&\times\frac{1}{n+1-|m|}
\left(\frac{n-|m|}{n+1-|m|}\right)
\power{\frac{1}{2}(a^{\dagger}+\mu b^{\dagger})(a+\mu^* b)},
\nonumber\\&&
\end{eqnarray}
where $\mu=m/|m|$.
Therefore, we have $G^{T_{a}}\geq0$ only if $n\geq|m|$, so that
$G$ is not separable for $n<|m|$. 
In view of (\ref{cond:Prepr}), then, the partial transpose is positive
whenever the Gaussian in question is $P$-representable.
And, as already remarked after (\ref{cond:Prepr}), $G$ is separable if it is
$P$-representable.
Together these observations say this:
\begin{equation}
  \label{peres+gauss}
  \parbox[b]{0.75\columnwidth}{%
   A positive Gaussian operator is separable 
   if it is $P$-representable, and only then.  }
\end{equation}
This statement is more generally true than our argument suggests,
because the limitations that originate in the special form of $\mathbf{V}$ of
(\ref{thisV}) can be lifted \cite{tutorial}.

Note, in particular, that the projectors of Sec.\ \ref{sec:pure} are
non-separable for $|m|=\sqrt{n(n+1)}>0$ which includes the EPR limit of
$n\to\infty$. 
The Gaussian projector (\ref{pureG}) is separable only in the other limit of
$m=0$, $n=0$, when it projects onto the two-oscillator ground state
$|0,0\rangle$.

\section*{Acknowledgments}
K.~W.\ would like to thank Professor H. Walther for the hospitality extended
to him in Garching.
This work was partially supported by a KBN Grant.

\ifpreprintsty
\begin{figure}
\caption{\label{TheFig}%
Concerning the parameters of the Gaussian operator associated with the
$\mathbf{V}$ matrix of (\ref{thisV}).
Only $n,|m|$ values above the dashed line are allowed by constraint
(\ref{constraint0}). 
According to (\ref{constraint1}), values on or above the solid line specify
positive Gaussians of the form (\ref{pureG}). 
For values on the solid line, the Gaussian operator is a projector.
Separable Gaussians belong to values on or above the dash-dotted line;
see (\ref{cond:Prepr}).}
\end{figure}

\newpage\thispagestyle{empty}
\centerline{\epsfig{file=gaussBR.eps}} 
\vfill
\centerline{\footnotesize Englert\&W\'odkiewicz, Fig.~\ref{TheFig}}
\fi


\begin{references}
\bibitem{epr}
A. Einstein, B. Podolsky, and N. Rosen, Phys.\ Rev.\ \textbf{47}, 777 (1935).
\bibitem{kbkw} 
K. Banaszek and K. W\'odkiewicz, \pra \textbf{58}, 4345 (1998).
\bibitem{fsbfkp}
A. Furusawa, J. L. S\o{}rensen, S. L. Braunstein, C. A. Fuchs, H. J. Kimble,
and E. Polzik, Science \textbf{282}, 706 (1998).
\bibitem{opkp}
Z. Y. Ou, S. F. Pereira, H. J. Kimble, and K. C. Peng, 
\prl \textbf{68}, 3663 (1992).
\bibitem{kimble} 
S. L. Braunstein, C. A. Fuchs, H. J. Kimble, and P. van Loock, 
LANL preprint quant-ph/0012001 (2000).
\bibitem{zoller}  
L.-M. Duan, G. Giedke, J. I. Cirac, and P. Zoller,
\prl \textbf{84}, 2722 (2000).
\bibitem{simon} 
R. Simon, \prl \textbf{84}, 2726 (2000).
\bibitem{werner} 
R. F. Werner, \pra \textbf{40}, 4277 (1989).
\bibitem{tutorial}
B.-G. Englert and K. W\'odkiewicz (unpublished).
\bibitem{peres} 
A. Peres, \prl \textbf{77}, 1413 (1996).
\bibitem{wigner}
E. Wigner, Phys.\ Rev.\ \textbf{40}, 749 (1932).
\bibitem{cahglau}
K. E. Cahill and R. J. Glauber, Phys.\ Rev.\ \textbf{177} 1857,1882 (1969).
\bibitem{wig+par}
Perhaps the first to note the intimate connection between the Wigner function
and the parity operator was A. Royer, \pra \textbf{15}, 449 (1977);
in the equivalent language of the Weyl quantization scheme the analogous
observation was made a bit earlier by A. Grossmann, 
Commun.\ Math.\ Phys.\ \textbf{48}, 191 (1976). 
A systematic study from the viewpoint of operator bases is given by 
B.-G. Englert, J. Phys.\ A: Math.\ Gen.\  \textbf{22}, 625 (1989).
\bibitem{schwinger}
J. Schwinger, in \textit{Quantum Theory of Angular Momentum}, 
edited by L. C. Biederharn and H. Van Dam (Academic, New York, 1965).
\bibitem{horodeccy} 
M. Horodecki, P. Horodecki, and R. Horodecki, \pl\textbf{A223}, 1 (1996).
\end{references}
\end{document}